\documentclass[journal]{IEEEtran}
\usepackage{amsmath}
\usepackage{amsfonts}
\usepackage{epsfig}
\usepackage{amssymb}
\usepackage{bbm}
\usepackage{graphicx}
\usepackage{float}
\usepackage{rotating}
\usepackage{color}
\usepackage{bbding}
\usepackage{textcomp}
\usepackage{multirow}
\usepackage[linesnumbered,ruled]{algorithm2e}
\usepackage[switch]{lineno}
\usepackage{amsthm}
\usepackage{amsopn}
\usepackage{url}
\usepackage{epstopdf}
\usepackage{color}
\usepackage{bm}
\usepackage{verbatim}
\usepackage{enumerate}
\usepackage [short,nocomma]{optidef}
\usepackage{mathtools}
\usepackage{booktabs}
\usepackage[caption=false]{subfig}
\usepackage{array}
\usepackage{pifont}
\usepackage[linesnumbered,ruled]{algorithm2e}
\usepackage{pifont}
\usepackage{booktabs}
\usepackage{multirow}
\newcommand{\cmark}{\ding{51}}
\newcommand{\xmark}{\ding{55}}

\usepackage{tikz}
\usepackage{wasysym}
\usepackage{threeparttable}
\usepackage{array}
\usepackage{multirow}
\usepackage{longtable}
\usepackage{rotating}

\SetKwInput{KwInput}{Input}
\SetKwInput{KwOutput}{Output}

\DeclareMathOperator*{\argmax}{arg\,max}

\ifCLASSINFOpdf
 
\else
 
\fi
\begin{document}

\title{Optimizing Age of Trust and Throughput in Multi-Hop UAV-Aided IoT Networks}
\author{Yizhou Luo, Kwan-Wu Chin, Ruyi Guan, Xi Xiao,  Caimeng Wang, Jingyin Feng, and Tengjiao He

}

\maketitle

\begin{abstract}
Devices operating in Internet of Things (IoT) networks may be deployed across vast geographical areas and interconnected via multi-hop communications.  Further, they may be unguarded.  This makes them vulnerable to attacks and motivates operators to check on devices frequently. 
To this end, we propose and study an Unmanned Aerial Vehicle (UAV)-aided attestation framework for use in IoT networks with a charging station powered by solar. 
A key challenge is optimizing the trajectory of the UAV to ensure it attests as many devices as possible.  A trade-off here is that devices being checked by the UAV are offline, which affects the amount of data delivered to a gateway.  Another challenge is that the charging station experiences time-varying energy arrivals, which in turn affect the flight duration and charging schedule of the UAV.
To address these challenges, we employ a Deep Reinforcement Learning (DRL) solution to optimize the UAV's charging schedule and the selection of devices to be attested during each flight.
The simulation results show that our solution reduces the average age of trust by 88\% and throughput loss due to attestation by 30\%. 
\end{abstract}

%

\IEEEpeerreviewmaketitle
%

\section{Introduction}
%
%
Internet of Things (IoT) networks may span vast geographical areas, such as farmlands, and they are used for livestock monitoring and to support dam/river management systems to name a few.
A key issue is device security.  In particular, devices are often unattended and lack physical security, meaning they are vulnerable to malicious attacks. 
Further, as they are resource-limited, devices are unable to run sophisticated security protocols, meaning they are easily compromised by attackers.
In this respect, an attacker may compromise trusted applications on a device, and/or manipulate their collected data.
Devices may be commandeered to facilitate data exfiltration or inject false data, or act as part of botnets to carry out Distributed Denial-of-Service (DDoS) attacks~\cite{6198335}.
Thus, there is a critical need to develop mechanisms to verify devices to ensure their integrity, e.g., they do not host malware.

One such mechanism is software attestation; this is in contrast to hardware-based attestation, e.g., \cite{8027075}, that is often costly, and not easily supported by existing devices as it requires hardware expansion and modifications~\cite{9096052}. 
Briefly, a software-based attestation mechanism involves a trusted verifier that aims to determine the integrity of a device (or prover).  A device is given a challenge whereby they conduct a checksum of their memory, which it uses in its attestation response.  A verifier then checks whether the response is valid; i.e., the checksum computed by a device matches the checksum at the verifier.
Further, a verifier may require the said response to be received by a given deadline.   This prevents a malware from hiding its presence when a device is checking its memory~\cite{1301329}.

Software-based attestation mechanisms, however, have two limitations:
i) {\em single-hop dependency}: traditional attestation protocols, e.g.,~\cite{1301329}\cite{10.1145/2508859.2516650}, require direct point-to-point connectivity between a verifier and a prover~\cite{9096052}.
They are, however, not suitable for multi-hop IoT networks as they are vulnerable to man-in-the-middle attacks.
In this respect, researchers have proposed swarm-based attestation methods, e.g., \cite{SEDA}.  However, they suffer from high signaling overheads, and require coordination from multiple devices, which is impractical when devices have resource constraints,
ii) {\em timing constraint}: as mentioned,  a verifier expects a response to a challenge to arrive within a given time.  This time constraint may not be satisfied in IoT networks whereby devices have a low duty cycle or when they experience energy outage; both of which delay the transmission of attestation responses.
\begin{figure}[t]
	\centering 
	\includegraphics[scale = 0.7]{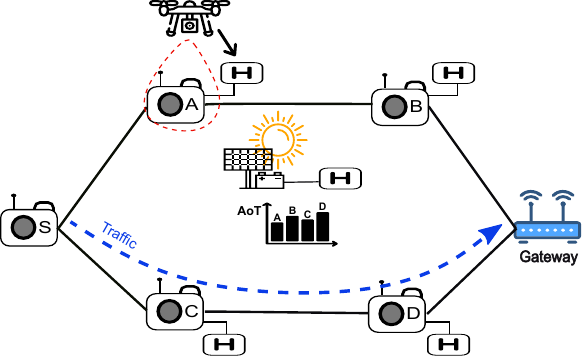}
	\caption{An example IoT network with a charging station powered by solar.  The UAV visits device A, B, C and D to check for malware.  When a device is being checked, where the shown UAV {\em lands} on the corresponding platform marked as `H' of node-A and connects to it via a {\em wired} or USB link, node-A does not forward traffic.  Consequently, device S has to use an alternative path to transmit its data to the gateway.}
	\label{ToyEG}
\end{figure}
To overcome the said limitations, we propose using an Unmanned Aerial Vehicle (UAV) as a verifier, with a key advantage being that devices do not require a hardware-based attestation module (e.g., Trusted Platform Module, TPM)~\cite{TPMbook}. During attestation, the UAV {\em lands} on a device, and thus they can establish a wired physical connection, e.g., via a Universal Serial Bus (USB).  This avoids any security and wireless issues, e.g., eavesdropping, interference, obstructions.   Further, using a wired link avoids intermediate nodes or wireless relays, and thus guarantees strict timing constraints and reliable single-hop data transfer.   
We also point out that the software attestation process may take a few minutes to complete.
Fig.~\ref{ToyEG} shows an example IoT network where attestation is carried out by a UAV, which in turn is supported by a charging station powered by solar.   In addition, devices forward data across multiple hops to the gateway.   Given this network, our aim is to determine the trajectory and charging schedule of the UAV as well as the routing of data.  

%
The proposed UAV-based software attestation solution addresses a number of challenges. 
First, an operator does not know when a device will be compromised by an attacker.  Consequently, an operator has to schedule a UAV to check on devices frequently.  A key challenge is that the UAV has limited battery lifetime, which affects its traveling distance or the number of devices it can visit in each trajectory.
Second, when a device is being checked, due to resource constraints, it will be unable to carry out normal functions, e.g., sampling and forwarding of traffic.  Consequently, the amount of data arriving at a gateway may reduce, which in turn reduces the coverage quality of one or more targets.
Third, as mentioned, the UAV is supported by a charging station, which is powered by a renewable energy source, meaning its available energy exhibits spatio-temporal properties.  This in turn affects the UAV's trajectory and how often it attests devices.
 
%
In this paper, we address the aforementioned challenges as follows. 
First, we employ the concept of Age of Trust (AoT)~\cite{xiao2024AoT}, which is defined as the elapse time since the last verification of a device.  
Second, to deal with stochastic energy arrivals at the charging station, we employ a reinforcement learning solution to optimize the trajectory of the UAV and also routing of traffic.
In a nutshell, this paper makes the following contributions:
\begin{itemize}
\item This paper is the first to study a novel UAV-based attestation system that leverages the said AoT metric, and reinforcement learning to ensure device integrity and throughput of a multi-hop IoT network supported by a charging station.
\item It is the first to outline a Markov Decision Process (MDP) that considers the spatio-temporal energy arrivals at the charging station, the UAV's location and the AoT of devices.
To determine the optimal policy that yields minimum average AoT and maximum network throughput, we outline a deep reinforcement learning (DRL) solution to determine both UAV's device verification trajectory and charge scheduling. 
\item It reports the first study of the problem and solution.  The simulation results show that after training, our DRL-based approach reduces the average AoT by 88\% and the network throughput loss due to attestation by 30\%. 
\end{itemize}
%

This paper is organized as follows. Section~\ref{sec:RWorks} highlights gaps in prior works. Section~\ref{sec:sysM} presents the system model, which includes the network architecture and operational constraints of interest. After that, it presents the optimization problem at hand. Section~\ref{sec:sol} introduces our DRL solution. Section~\ref{sec:eval} presents our evaluation methodology and results. Finally, Section~\ref{sec:conc} presents the conclusions and future works.

%
%
\section{Related Works}\label{sec:RWorks}

\begin{table*}[htbp]
    \label{tab:uav_comparison}
    \caption{Comparison of prior works on age-optimization in UAV-assisted IoT Networks.}
    \centering
    \renewcommand{\arraystretch}{1.2}
    \centering
    \renewcommand{\arraystretch}{1.2}
    \resizebox{\textwidth}{!}{
    \begin{tabular}{lccccccccc}
        \toprule
        \textbf{Citation(s)} & \textbf{Age} & \textbf{UAV} & \textbf{Energy} & \textbf{Battery} & \textbf{Random} & \textbf{Renewable} & \textbf{Multi-hop} & \textbf{Traffic} & \textbf{Learning} \\
        \textbf{} & \textbf{Metric} & \textbf{Charging} & \textbf{Efficiency} & \textbf{Management} & \textbf{Energy} & \textbf{Energy} & \textbf{Network} & \textbf{Flow} & \textbf{Algorithm} \\
        \midrule
        ~\cite{10869306, 10820863} & AoI& \xmark & \xmark & \xmark & \xmark & \xmark & \xmark & \xmark & \cmark \\
        ~\cite{10836723} &  AoI& \cmark & \xmark & \cmark & \xmark & \xmark & \xmark & \xmark & \cmark \\
        ~\cite{10877810} &  AoI& \cmark & \cmark & \cmark & \xmark & \xmark & \xmark & \xmark & \cmark \\
        ~\cite{10838612} & AoC & \cmark & \cmark & \cmark & \xmark & \xmark & \xmark & \xmark & \cmark \\
        ~\cite{10759639} & Age of Task & \xmark & \xmark & \cmark & \xmark & \xmark & \xmark & \xmark & \cmark \\
        ~\cite{10188891} & AoS & \xmark & \xmark & \xmark & \xmark & \xmark & \xmark & \xmark & \cmark \\
        ~\cite{9562134} & AoU & \xmark & \xmark & \xmark & \xmark & \xmark & \xmark & \xmark & \cmark \\
        ~\cite{10858287} & AoV & \xmark & \xmark & \xmark & \xmark & \xmark & \xmark & \xmark & \cmark \\
        {\bf This study} & Age of Trust & \cmark & \cmark & \cmark & \cmark & \cmark & \cmark & \cmark & \cmark \\
        \bottomrule
    \end{tabular}}
\end{table*}
A number of works have focused on \emph{age}-optimization problems in UAV-assisted networks, where UAVs act as mobile data collectors.  A popular metric is the Age of Information (AoI), which is defined as the time elapsed since the most recently received data packet was generated at the source. Several studies have explored UAV-based data collection frameworks that aim to minimize AoI while considering system constraints such as UAV energy consumption, mobility, and network dynamics. 
For example, reference~\cite{10877810} proposes a Deep Q-Network (DQN)-based algorithm to optimize UAV flight paths and data upload intervals while reducing collision risks. The study incorporates UAV energy constraints, kinematic limitations, and the mobility of sensor nodes, with the objective of maximizing data uploads and energy efficiency while minimizing AoI.
Reference~\cite{10869306} considers dynamic data upload demands, network complexity, and uncertain data generation, aiming to optimize data rate, AoI, and UAV energy consumption simultaneously.
Reference~\cite{10836723} studies a system with UAVs, sensor nodes, workers, and a data center.  
It outlines DRL-based optimization framework to determine UAV flight paths and assignments of tasks to workers to minimize global AoI.
Reference~\cite{10820863} investigates an Space–Air–Ground Integrated Network (SAGIN) that includes satellites, high-altitude platforms (HAPs), UAVs, and IoT devices. The study applies a DRL-based algorithm to determine UAV and HAP placement, UAV trajectories, and UAV-IoT device matching to achieve minimum AoI while addressing network construction costs and real-time data transmission constraints.

Beyond AoI, researchers have introduced new age-related metrics to evaluate system performance in UAV-assisted networks.
For example, reference~\cite{10838612} introduces the concept of Age of Charge (AoC) to represent the elapse time since an IoT device was last charged. The authors consider a system where a UAV travels among IoT devices and charges them wirelessly. 
The authors apply Deep Deterministic Policy Gradient (DDPG) to determine the UAV’s path, velocity, and charging schedule to maximize the total energy harvested by IoT devices whilst minimizing the UAV’s energy consumption.
The UAVs in~\cite{9562134} use a distributed online learning solution to collaboratively train Artificial Intelligence (AI) models.  They use Age of Update (AoU) to evaluate data freshness and training contributions, with UAVs being selected for training based on their local dataset size. 
In reference~\cite{10188891}, an Age of Synchronization (AoS) metric is used to measure both freshness and content of packets. 
To minimize AoS, the authors propose a DQN based algorithm to jointly optimize their UAV's flight trajectory and the upload scheduling policy of sensor nodes.
The work in~\cite{10759639} introduces the concept of the age of task, defined as the time interval from when a task data packet is generated to when the response for the task is returned. Each ground node generates emergency computation tasks with certain packet size and CPU cycles requirement. 
A UAV functions as an edge mobile server that collects and processes tasks. The authors aim to minimize the long term average weighted age of task by optimizing the UAV's trajectory, computation resource, transmit power of ground nodes and their task offloading decisions.
The authors of~\cite{10858287} introduce Age of Verification (AoV), which they use to verify generative AI models. UAVs travel to edge servers to verify model validity, ensuring that user tasks do not rely on outdated or mismatched models. The study employs zero-knowledge proof methods to prevent information leakage and then applies a multi-agent reinforcement learning algorithm to optimize UAV trajectories and verification strategies.
Although there are many works that have considered age related metrics in UAV-aided communication systems, no work has studied AoT optimization in IoT networks and proposed a UAV-aided solution.
Table~\ref{tab:uav_comparison} compares existing works in terms of age metric, object, battery management, energy status, network architecture, traffic, and UAV charging. 
We jointly consider UAV charge scheduling, energy management, time-varying energy arrivals, traffic flow when optimizing AoT in multi-hop IoT networks. 
Works such as~\cite{10869306}\cite{10820863}\cite{10858287} do not consider age optimization under limited resources, where they usually overlook the battery life of a UAV. 
They aim to optimize the UAV's energy efficiency given unlimited energy resource and do not consider UAV charging.
Furthermore, we consider scenarios in which systems are powered by renewable energy with spatio-temporal properties. 
This means UAV will receive varying amount of energy over time, which in turn affects its trajectory or the number of devices it can check.
Lastly, unlike past works, we consider the impact of attestation on network performance, where in our case, a device is offline when it is being checked by a UAV.  
%

%
\section{System Model}\label{sec:sysM}
Time is divided into discrete intervals, indexed by $ t \in \mathcal{T}=\{ 1, 2, \dots, T \}$.  Each time slot has a duration of $\tau$ seconds; in practice, attestation of a device's memory content may take hundreds of seconds or a few minutes. In this respect, $\tau$ can be set to the maximum attestation time.  We also note that our work continues to apply if devices have different resources or/and security requirements.
For ease of exposition, we consider traffic from a source node $s$ to a destination $d$. 
There is one UAV, denoted as $u$.
We model the network as a directed graph $G(\mathcal{V} \cup \{s\} \cup \{d\}, \mathcal{E})$, where $\mathcal{V}$ is the set of ground devices, and $\mathcal{E}$ contains communication links. There are $|\mathcal{V}|$ ground devices, where $|.|$ denotes set cardinality.
These devices are randomly distributed on an area with size $K \times K$ square meters. 
The coordinate of node $i$ is $(x_{i},y_{i})$. 
Base location $o$ is located at coordinate $(x_{o},y_{o})$. 
Define an indicator $o^t$ to represent whether UAV $u$ is at the base location ($o^t=1$) or not ($o^t=0$) in slot $t$.
Further, the base location has a solar-powered charging station; see details in Section~\ref{EH_model}.

The UAV is equipped with a hardware fingerprinting module, e.g.,~\cite{swA2016}.
It travels to the location of each node in $\mathcal{V}$ at the start of a time slot.  It {\em lands} on the node and connects to the node via a wire, e.g., USB. This connection enables reliable, low-latency verification of the node's memory contents.  We {\em emphasize} that we do not consider a wireless channel because a wired channel avoids issues such as eavesdropping, jamming, fading, and poor channel conditions.
The said verification process is assumed to take one time slot.
%

%
\subsection{Network Flow}
Define the maximum flow or throughput from source $s$ to destination $d$ as $R(G)$.    Note, in our work, we do not aim to optimize $R(G)$.  Instead, we assume there exist methods that can be used to compute $R(G)$.
More specifically, if links operate on the same channel, then the link scheduling used by the network will have an impact on $R(G)$.  In this respect, methods such as \cite{JointRLinkS} can be used to jointly optimize routing and link scheduling to maximize $R(G)$.  Apart from that, if links operate on an orthogonal channel, this problem can be solved using methods such as \cite{ChannelRLinkS}.  
When a device is being checked, it stops forwarding traffic; this is reasonable as devices are resource-limited and it also stops a malware from transmitting itself to another device.
Define an auxiliary variable $z^t_i\in\{0,1\}$ to represent whether device $i$ is chosen by UAV $u$ for attestation at time $t$; if it is chosen, then we have $z^t_i=1$; otherwise, $z^t_i=0$.    
Further, as UAV $u$ checks at most one device at a time, we have
\begin{equation}
   \sum_{i\in\mathcal{V}} z^t_i \le 1, \forall t\in\mathcal{T}.
\end{equation}

If device $i$ is selected for attestation ($z^t_i=1$), then we have a new graph $G'[i] = G(\mathcal{V} \setminus \{i\}, \mathcal{E} \setminus \mathcal{E}_i )$; this graph omits node $i$ and its incident edges.  We can then apply methods such as \cite{JointRLinkS, ChannelRLinkS} to compute $R(G'[i])$.
We also define
\begin{equation}
  R^t =
  \begin{cases}
   R(G'[i]) & z^t_i=1 \wedge \exists z^t_i \neq 0, \\
   R(G) & Otherwise.
  \end{cases}
\end{equation}
\subsection{Flight Model}
The travel distance $D_{ij}$ between any two nodes $i$ and $j$ is calculated by their Euclidean distance as
\begin{equation}\label{travel_distance}
D_{ij} = \sqrt{(x_i - x_j)^2 + (y_i - y_j)^2}.
\end{equation}
The UAV operates at a fixed altitude $H$. 
It takes $\tau$ seconds to fly between node $i$ to $j$. 
Let $V_{i j}$ be the UAV's flight speed between node $i$ and $j$, 
which is calculated as 
\begin{equation}\label{flight_speed}
V_{i j} = \frac{D_{ij}}{\tau}.
\end{equation}
\subsection{UAV Power Model}
The energy consumption of a UAV is only determined by its travel distance and flight speed.
The energy consumption $e_{i j}$ of the UAV when it moves between coordinate $(x_i, y_i)$ and $(x_j, y_j)$ is given by $e_{i j} = P_{i j} \cdot D_{i j}$, where $P_{i j}$ represents the UAV's energy consumption rate per meter traveled. 
According to the model in~\cite{8663615}, the energy consumption rate $P_{i j}$ (in Joule per meter) is a function of a UAV's flight speed $V_{i j}$, which is calculated as
\begin{align*}\label{energy_consumption_model} 
P_{i,j}=&P_{0} \left({ \frac {1}{V_{i,j}{}} + \frac {3V_{i,j}}{{V_{0}} ^{2}}}\right) \\&{}+ P_{1} \left({\sqrt {{V_{i,j}}^{-4} + \frac {1}{4{V_{1}}^{4}}}-\frac {1}{2{V_{1}}^{2}}}\right)^{1/2} \\&{}+ \frac {1}{2}\kappa D_{\mathrm{air}} S_{rot} A_{\mathrm{rot}} {V_{i,j}}^{2}\quad \forall i,j \in \mathcal{V}, \tag{5}\end{align*}
where $V_{0}$ is the tip speed of the rotor blade,  $V_{1}$ is the average hovering rotor induced velocity, $\kappa$ is the fuselage drag ratio, $S_{rot}$ is the rotor solidity and $A_{\mathrm{rot}}$ is the rotor disc area, $P_{0}$ and $P_{1}$ are the hovering blade profile power and induced power, and $D_{\mathrm{air}}$ is the air density. 
\subsection{UAV Energy Model}
The UAV's maximum energy capacity is $B_{\text{max}}$. Its energy level at time slot $t$ is  $b_u^t$. 
Once its energy level $b_u^t$ drops below a specified safety margin $B_{\text{min}}$, it must return to the base location for recharging, which requires one time slot. The threshold $B_{\text{min}}$ is set to the total energy consumed to travel to the next location, and from the said location to the base location. 
This ensures the UAV always has sufficient energy to return to the base location. 
We assume the UAV consumes negligible amount of energy when verifying a node; this is reasonable as the verification  operation is carried out by nodes.  

UAV $u$'s energy level evolves as 
\setcounter{equation}{5} 
\begin{equation}\label{UAVEvol}
  b_u^t =
  \begin{cases}
   b_u^{t-1} - e_u^t & {o^t = 0,} \\
   \min(B_{\text{max}},\,b_u^{t-1} - e_u^t + \hat{e}_u^t) & {o^t = 1,} 
  \end{cases}
\end{equation}
where $e_u^t = e_{i j}$ is the energy consumed to travel from coordinate $(x_i, y_i)$ to $(x_j, y_j)$ in slot $t$, and $\hat{e}_u^t$ represents the amount of energy received at the base location. If the UAV is {\em not} located at the base location, $\hat{e}_u^t$ is zero. 
Note, Eq.~\ref{UAVEvol} can also consider the case where a UAV replenishes the energy of devices.  Recall that the UAV is connected to a device via a USB link. In this respect, the energy consumption rate of the UAV can also include the charging of a device after attestation. 
\subsection{Energy Harvesting Model}\label{EH_model}
The base location has a battery that can store $\hat{B}_\text{max}$ amount of energy. 
As mentioned, it is powered by solar energy, where the energy arrivals at the base location are modeled using the Markov process described in~\cite{7008488}. This model includes four distinct solar states: ``Excellent'',  ``Good'', ``Fair'' and ``Poor''; each representing a weather condition. In the $j$-th state, the amount of energy harvested, $\hat{x}$ (in mJ), is a random variable drawn from a Normal distribution, $\mathcal{N}(\hat{x}\,|\,\mu_j,\sigma_j)$, with mean $\mu_j$ and variance $\sigma_j$. 
The energy harvested by the base location during time slot $t$ is given by $\hat{E}_o^t =  \Phi \cdot \overline{\eta} \cdot \tau \cdot \hat{x}^t$, where $\Phi$ is the size of the solar panel and $\overline{\eta}$ is the solar energy conversion efficiency~\cite{7008488}.  The energy level $\hat{B}_o^t$ at the base location evolves as
\begin{equation}
    \hat{B}_o^t=    \min(\hat{B}_\text{max},\,\hat{B}_o^{t-1}-o^t\hat{e}_u^t+\hat{E}_o^t),
\end{equation}
where $\hat{e}_u^t$ (in mJ) represents the amount of energy used to recharge a UAV at time slot $t$, which is limited as per $\hat{B}_o^t \geq \hat{e}_u^t$.  
%

%
\section{Problem Formulation}\label{sec:prob}
We adopt the AoT model~\cite{xiao2024AoT} to quantify the recency in which a device has had an integrity check.
Each device $i \in V$ is associated with an AoT value $\delta^t_i$.  The value $\delta^t_i$ increases by one after each time slot if a device is not verified.  Once verified, the AoT value of a device $\delta^t_i$ is reset to one. Formally, its AoT evolves as 
\begin{equation}
  \delta^t_i =
  \begin{cases}
   \delta^{t-1}_i + 1 & {z^t_i = 0,} \\
   1 & {z^t_i = 1.} 
  \end{cases}
\end{equation}
Given a time horizon with $T$ time slots, our problem involves optimizing two objectives: i) minimizing the average AoT of all devices.  Formally, 
\begin{equation}\label{avgaot}
\overline{\delta}^t = \sum_{i = 1}^{|\mathcal{V}|} \frac{\delta_i^t}{|\mathcal{V}|},
\end{equation}
and ii) maximizing the {\em average} flow between the source and destination nodes, $s$ and $d$. Define the following weighted objective function:
\begin{equation}\label{policy2}
Z(\pi)=\lim_{T\rightarrow\infty} \mathbb{E}^{\pi} \left[ \sum_{t=1}^T \theta_1 R^t - \theta_2 \overline{\delta}^t \right],
\end{equation}
where $\theta_1, \theta_2 \ge 0$ are weights that balance the flow of a network and the AoT of devices, and $\mathbb{E}^{\pi}[.]$ denotes the expectation when using policy $\pi$. 

Note that frequent device verifications will decrease the amount of data flowing through the network. Therefore, there is a trade-off between the flow of a network and the AoT of devices. 
In addition, the UAV may not obtain sufficient energy to travel between devices. The UAV needs to make the best use of limited energy to reduce the AoT of devices.
Let $\Omega$ be a collection of policies, where a policy $\pi\in\Omega$ determines the UAV's destination node $i$ at each time slot $t$. Note, there are $|\mathcal{V}|$ devices and one base location, so we have $i \in \{1,\dots,|\mathcal{V}|+1\}$. Our aim is to find the best policy $\pi^*$ that maximizes the flow of a network while minimizing the AoT of devices. Formally, 
\begin{equation}\label{obj}
\pi^* = \argmax_{\pi} Z(\pi).
\end{equation}

\section{Solution}\label{sec:sol}
In this section, we first formulate the problem in Section~\ref{sec:prob} as an MDP and then introduce a Prioritized Dueling Double Deep Q-network (PD3QN)-based solution.
\subsection{Markov Decision Process (MDP)}
An MDP~\cite{BELLMAN1957MDP} consists of three key elements:
\subsubsection{State space $\mathcal{S}$}  It represents the states of an environment. The state at time $t$ is denoted as $s_t$. 
It consists of (i) the AoT $\delta^t_i$ of each device, (ii) the current position $p^t$ and battery level $b_u^t$ of the UAV, and (iii) the energy reserve $\hat{B}_o^t$ of the base location. Formally, we have
\begin{equation}\label{state}
s_t = \{ \delta^t_1, \delta^t_2 \dots,\delta^t_{|\mathcal{V}|},p^t,b_u^t,\hat{B}_o^t \}.
\end{equation}
Note that UAV $u$ is either located at a node or the base location. Hence, $p_t \in \{1,2,\dots,|\mathcal{V}|+1 \}$ is an integer; note, $p_t = |\mathcal{V}|+1$ corresponds to the base location.
\subsubsection{Action space $\mathcal{A}$}  It represents all possible decisions afforded to an agent.  Let $a_t$ be the action taken at time $t$, where in our case, we have action $a_t = i$, meaning the UAV will fly to node $i$ in the next time slot. Note, there are $|\mathcal{V}|$ ground devices and a base location. Hence, $|\mathcal{A}|$ is equal to $|\mathcal{V}|+1$ and $a_t = |\mathcal{V}|+1$, which correspond to the UAV returning to the base location to recharge its battery.

\subsubsection{Reward $r_t(s_{t+1}|s_t,a_t)$} It represents the immediate reward obtained by an agent after taking action $a_t$ in state $s_t$. 
In our case, it is defined as $r_t(\cdot) = r^t_1 + r^t_2$, where $r^t_1 = \theta_1 \cdot R^t$.
The term $r^t_2 = \theta_2 \cdot(\overline{\delta}^{t+1} - \overline{\delta}^t)$ indicates that if the average AoT of devices in the next slot $t+1$ is greater than that of the current slot $t$, the reward $r^t_2$ is negative, and vice-versa. 
The weights $\theta_1,\theta_2$ allow an operator to balance between network throughput and the AoT of devices.

The goal is to learn the optimal policy $\pi^*$, which yields the optimal action for each state that maximizes the following expected cumulative discounted reward:
\begin{equation}\label{long-term accumulative reward}
\mathop{\mathbb{E}}\left[\sum_{t=0}^{+\infty}\gamma^t r_t(s_{t+1}|s_t,\pi(s_t))\right],
\end{equation}
where $\gamma \in [0,1]$ is the reward discount factor that balances the importance of immediate and future rewards. The aforementioned MDP satisfies the Markov property. This is because after taking an action, the UAV will travel to another position, which moves the current state to the next state. 

Given a policy $\pi$, the Q-value $Q^{\pi}(s_t, a_t)$ is defined as the expected sum of discounted future rewards when the agent starts from state $s_t$, takes action $a_t$, transits to state $s_{t+1}$, and thereafter follows policy $\pi$. Mathematically,
\begin{equation}\label{Q-value}
Q^{\pi}(s_t, a_t) = \mathop{\mathbb{E}}\left[\sum_{t=0}^{+\infty}\gamma^t r_t(s_{t+1}|s_t,a_t=\pi(a_t))\right].
\end{equation}
In other words, $Q^\pi(s, a)$ quantifies how good it is to take action $a_t$ in state $s_t$ and then continue following $\pi$.

To maximize rewards, we seek the optimal policy $\pi^*$ that leads to the maximum Q-value, denoted by $Q^*(s, a) = \max_{\pi} \, Q^{\pi}(s, a)$. A common algorithm to learn $Q^*(s, a)$ from experience, i.e., by interacting with the environment, is Q-learning. The Q-learning update rule is given by the following Bellman expectation equation:
\begin{equation}\label{q-learning}
Q(s, a) \leftarrow Q(s_t, a_t)
+ \eta \Big[
\underbrace{r_t + \gamma \max_{a' \in \mathcal{A}} Q(s_{t+1}, a')}_{\text{Target Q-value}}
- Q(s, a)
\Big],
\end{equation}
where $\eta \in (0, 1] $ is the learning rate. The action $a'$ is the expected action with the maximum Q-value in $s_{t+1}$.

\subsection{PD3QN}
Prioritized Dueling Double Deep Q-network (PD3QN) is a state-of-the-art reinforcement learning algorithm designed to address key challenges in the Q-learning framework. These challenges include the curse of dimensionality, overestimation bias, inefficient state-action value estimation, and inefficient experience replay. PD3QN integrates three key enhancements to the Deep Q-Network (DQN): Double Q-Networks, Dueling Architecture, and Prioritized Experience Replay (PER). Next, we detail each key component and their integration within the PD3QN framework.

\subsubsection{Double Q-Networks}
Traditional Q-learning algorithms store Q-values in a Q-table and iteratively update them. 
However, for high-dimensional or continuous state spaces, such as in our problem where the state space includes continuous variables, e.g., battery level $b_u^t$ and $\hat{B}_o^t$, and AoT value of nodes $\delta^t_i$, the Q-table approach becomes computationally intractable due to the exponential growth in state-action pairs.
To address this, Q-learning is often combined with a neural network to approximate the Q-value function, commonly referred to as a Q-Network~\cite{Mnih2015HumanlevelCT}.

When using neural networks for Q-value approximation, the overestimation of target Q-values can occur. 
This is because the same Q-Network is used for both selecting $a'$ via the max operator and for evaluating the Q-value of selected actions $Q(s_{t+1}, a')$, see Eq.~\eqref{q-learning}.
To mitigate overestimation, Double Deep Q-Networks (DDQN) decouple the action selection and evaluation processes using two networks: i) An evaluation network with parameters $w$ outputs evaluated Q-values $Q(s,a;w)$ and selects an action $a^*$ with the maximum Q-value, and ii) A target network with parameters $w'$ is used to evaluate the action $a^*$ selected by the evaluation Q-Network $w$~\cite{double_Q}.
Concretely, the target value $\hat{Q}(\cdot)$ for DDQN is given by
\begin{equation}\label{DDQN_target}
\hat{Q}(s_{t+1},a^*,r_t;w')=r_t+\gamma Q(s_{t+1},a^*;w'),
\end{equation}
where $a^* = \arg\max_{a'} Q(s_{t+1},a';w)$. This decoupling ensures that the action evaluation does not suffer from the same overestimation bias as the action selection, leading to more accurate Q-value estimates.

\subsubsection{Dueling Architecture}
In many states, action choices have minimal impact on rewards (non-critical states), yet standard Q-networks inefficiently allocate significant network capacity to estimate Q-values for all actions, leading to slow convergence. Conversely, identifying ``vulnerable'' or high-impact states is crucial, but conventional Q-learning lacks explicit mechanisms to prioritize state-action importance.

The dueling network architecture addresses this by factorizing the Q-function into two components~\cite{DuelingPaper}:
i) A single scalar $V(s_t)$ to quantify the overall ``goodness'' of state $s_t$. Formally,
\begin{equation}\label{dueling_value}
V^{\pi}(s_t) = \mathop{\mathbb{E}_{a_t \sim {\pi}(s_t)}}[Q^{\pi}(s_t, a_t)]. 
\end{equation}

ii) A vector $A(s_t,a_i)$ whose size is equal to the number of possible actions. Each component of $A(s_t,a_i)$ indicates how much better (or worse) action $a_i$ is as compared to others in the same state $s_t$. Formally,
\begin{equation}\label{dueling_advantage}
A^{\pi}(s_t, a_t)=Q^{\pi}(s_t, a_t)-V^{\pi}(s_t),
\end{equation}
with the constraint $\mathop{\mathbb{E}_{a_t \sim {\pi}(s_t)}}[A^{\pi}(s_t, a_t)] = 0$.
This decomposition enables agents to identify critical states and and prioritize actions.

\begin{figure}[htbp]
	\centering 
	\includegraphics[scale = 0.75]{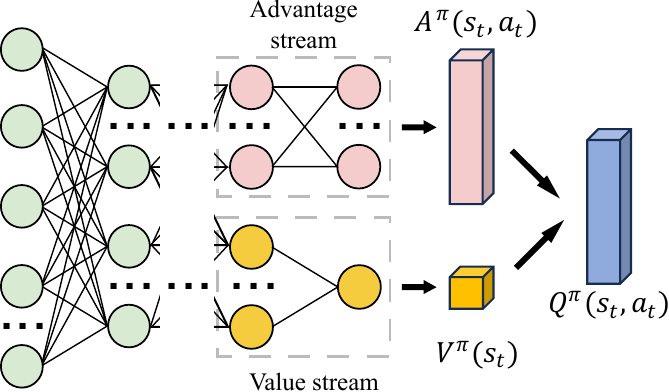}
	\caption{The dueling architecture. On the left side is a Q-network with its neurons (green circles). The Q-Network is then divided into two streams with their own neurons: the Advantage stream (pink circles) and the Value stream (yellow circles). The Advantage stream estimates $A^\pi(s_t, a_t)$, and the Value stream estimates $V^\pi(s_t)$. Finally, these two components are combined to obtain $Q^\pi(s_t, a_t)$.
    }
	\label{Dueling_architecture}
\end{figure}

As illustrated in Fig.~\ref{Dueling_architecture}, the dueling network architecture uses two parallel data streams to estimate $V(s_t; l, w)$ and $A(s_t, a_t; l', w)$, where $l$, $l'$ are stream weights and $w$ are overall parameters of a Q-network. The output layer aggregates the two streams to yield the Q-values:
\begin{equation}\label{Dueling Q-value}
Q(s_t, a_t; l, l', w) = V(s_t; l, w) + A(s_t, a_t; l', w) - \overline{A}(s_t; l', w), 
\end{equation}
where
\begin{equation}\label{}
\overline{A}(s_t; l', w) = \frac{1}{|\mathcal{A}|} \sum_{a' \in \mathcal{A}} A(s_t, a'; l', w). 
\end{equation}
As pointed out in~\cite{DuelingPaper}, when a Q-value is provided, it is impossible to uniquely determine its $V(\cdot)$ and $A(\cdot)$. By subtracting the average advantage $\overline{A}(\cdot)$, this issue can be resolved.
In addition, the zero-mean constraint $\mathop{\mathbb{E}_{a_t \sim {\pi}(s_t)}}[A^{\pi}(s_t, a_t)] = 0$ ensures the Q-network to learn distinct representations for $V(\cdot)$ and $A(\cdot)$.  

When choosing an action, we apply the $\epsilon$-greedy policy to balance the trade-off between exploration, where an agent tries out different actions to gather information and exploitation where an agent chooses the best-known action to maximize rewards~\cite{sutton2018RLBOOK}. Specifically, an agent choose $a_t$ according to
\begin{equation}\label{}
a_t =
\begin{cases}
\text{random action from } \mathcal{A} & \text{with probability } \epsilon_G, \\
\arg\max_{a \in \mathcal{A}(s_t)} Q(s_t, a) & \text{with probability } 1 - \epsilon_G.
\end{cases}
\end{equation}
where $\epsilon_G$ is the exploration rate and $0 \leq \epsilon_G \leq 1$.

\subsubsection{Training with Prioritized Experience Replay}
An agent interacts with an environment and collects experiences, typically stored as tuples $(s_t, a_t, r_t, s_{t+1})$ in a memory replay buffer $\mathcal{D}$. Formally, we have
\begin{equation}\label{}
(s_i, a_i, r_i, s'_i) \quad \text{for} \; i \in \{1,2,\dots,|\mathcal{D}|\}. 
\end{equation}
During training, the agent samples mini-batches $\mathcal{B} \subseteqq \mathcal{D}$ of these stored experiences to perform gradient-based updates of the Q-network.
However, this is inefficient because experiences are not equally useful~\cite{schaul2016prioritizedexperiencereplay}.
Some experiences, particularly those with a high “\textit{surprise}” or large temporal-difference (TD) error, are more informative for learning. For a transition $i$, the TD error $\hat{\delta_i}$ is computed as
\begin{equation}\label{}
\hat{\delta_i} = \bigl| \hat{Q}_i(s'_i, a^*_i, r_i;w') - Q(s_i, a_i ; w) \bigr|, 
\end{equation}
where $\hat{Q}_i(\cdot)$ is the target Q-values calculated as per~\eqref{DDQN_target}. Intuitively, if $\hat{\delta_i}$ is large, it suggests that our current Q-network has a large error, meaning it could be more beneficial to train on it more often to reduce the average error.

To this end, Prioritized Experience Replay (PER) modifies uniform sampling by giving more frequent sampling opportunities to those experiences with higher TD errors~\cite{schaul2016prioritizedexperiencereplay}.  
Each transition $i$ is assigned a priority $p_i$, defined as
\begin{equation}\label{}
p_i = |\hat{\delta}_i| + \epsilon_{P},
\end{equation}
where $\epsilon_{P} > 0$ is a small constant, i.e., priority bias, that avoids the possibility of zero priority, which would cause some transitions never to be sampled.

The replay buffer then samples transitions proportionally to their priorities. Concretely, the probability $\hat{P}(i)$ of sampling transition $i$ is given by
\begin{equation}\label{}
\hat{P}(i) = \frac{p_i^\alpha}{\sum_{k=1}^{|\mathcal{D}|} p_k^\alpha},
\end{equation}
where $\alpha \ge 0$ is a hyperparameter that controls how much prioritization is used. if $\alpha = 0$, it reduces to uniform sampling. A larger value of $\alpha$ make the agent focus more on transitions with higher priority.

Non-uniform sampling introduces bias, where transitions with higher priority are oversampled, while those with low priority are undersampled~\cite{schaul2016prioritizedexperiencereplay}.
This can be corrected by importance sampling weights $\hat{w}_i$, defined as
\begin{equation}\label{}
\hat{w}_i = \left(\frac{1}{|\mathcal{D}| \, \hat{P}(i)}\right)^{\beta} \Big/ \max_j \hat{w}_j 
\end{equation}
where $|\mathcal{D}|$ is the size of the memory replay buffer and $\beta$ is a hyperparameter that anneals from 0 up to 1 during training, controlling how much to compensate for the prioritized sampling distribution. When $\beta = 0$, no correction is applied. When $\beta = 1$, we fully compensate for the non-uniform sampling.
The denominator $\max_j \hat{w}_j$ normalizes the values to $[0,1]$.

With the above mechanism, the gradient updates to the Q-network incorporate the importance-sampling weights. Instead of Q-values maximization, we minimize the average weighted temporal difference-error of Q-values $\hat{\delta_i}$. During the training phase, we extract a mini-batch of samples $\mathcal{B}$ from $\mathcal{D}$ according to $P(i)$ at each time slot to refine its model, a process referred to as a single training step. The weighted loss function $L(w)$ is
\begin{align}\label{loss_func}
L(w) &= \sum_{i \in \mathcal{B}} \hat{w}_i \cdot (\hat{\delta_i})^2 \\
&= \sum_{i \in \mathcal{B}} \hat{w}_i \bigl(\hat{Q}_i(\cdot; w') - Q(\cdot; w)\bigr)^2.
\end{align}
Hence, high-TD-error transitions are sampled more often but will be down-weighted, whereas low-error transitions are sampled less often but will be up-weighted.
We apply soft update strategy during the training phase. The target Q-Network is updated less frequently to prevent large fluctuations in the target Q-values and reduce the risk of divergence during training. For every $U$ training steps, the parameters of the target network are updated as follows:
\begin{equation}\label{}
w_t' \leftarrow \hat{\tau} w_t + (1 - \hat{\tau}) w'_t,
\end{equation}
where $w_t$ and $w'_t$ are the parameters of the evaluation and target Q-networks at time $t$. The constant $\hat{\tau}$ is the soft update factor, where $0 < \hat{\tau} \ll 1$; this effectively blends the current online parameters $w$ with the existing target parameters $w'$.

%
%
\section{Evaluation} \label{sec:eval}
We studied our solution in Python 3.11 with PyTorch 2.5.1 on a workstation with Intel(R) Core(TM) i9-14900K, NVIDIA GeForce RTX 4090D and 64 GB of RAM. Unless otherwise stated, we use the parameter values listed in Table~\ref{simparameters}. We consider a network with seven ground devices randomly distributed in a 2.5 $\times$ 2.5 km square area. The UAV is a DJI Mavic 3.  Its battery capacity is 77 W·h and maximum velocity is 21 m/s. It starts from the base location. The base location is at $(0,0)$. 
The maximum battery capacity of the base location is $\hat{B}_\text{max} = 0.77$ kWh, with an initial energy of $0.385$ kWh.

A PD3QN agent is deployed on the UAV. 
It employs two identical dueling Q‑networks. Each network processes a 10‑dimensional state input through a 256‑unit dense hidden layer. Each network has value and advantage streams, each realized by its own 256‑unit dense hidden layer. The two streams are then recombined into an eight‑dimensional Q‑value output. We use the Rectified Linear Unit (ReLU) function as the activation function of neurons.
We use the Adam optimizer, i.e., adaptive moment estimation algorithm.

We also evaluated hyperparameter configurations within theoretically justified ranges, including learning rate $\eta$ from $10^{-2}$ to $10^{-5}$, decay rate $\gamma$ from 0.1 to 0.9, and priority exponent $\alpha$ from 0.1 to 0.5. Then, we choose best hyperparameter values for subsequent analyses, as detailed below. The learning rate $\eta$ and reward decay rate $\gamma$ is set to $10^{-4}$ and 0.5. We set the initial exploration rate to 1.0 and then gradually decrease it to 0.05. This enables the UAV to gather diverse experiences initially and then shifts to exploit its learned knowledge for reward maximization after convergence. The UAV has a memory buffer to store up to 4,000 samples. The priority exponent $\alpha$ and priority bias $\epsilon_{P}$ are set to 0.2 and 10$^{-5}$, respectively. Additionally, we adopt dynamic sampling exponent $\beta$, which starts at 0.6 and gradually anneals to 1.0 over time to correct for the sampling bias introduced by prioritization. 

Each experiment is run for 100 episodes; each episode has 2,000 time slots.  We use the initial 80 episodes to train the UAV, and then assess the UAV in the last 20 episodes. 
In each episode, we record the average value of the following metrics: 
\begin{itemize}
\item \textbf{Average reward}. This is calculated as $\frac{1}{2000}\sum_{t = 1}^{2000}{r_t}$, which is the average reward over one episode.
\item \textbf{Average AoT of sensor nodes}. This metric corresponds to the average AoT of sensor nodes.  Specifically, we calculate $\overline{\delta}^t$ as per Eq.~\eqref{avgaot}. Its average over one episode is calculated as $\frac{1}{2000}\sum_{t = 1}^{2000}{\overline{\delta}^t}$.
\item \textbf{Network throughput} in Kilobit per second (Kbps).  This metric corresponds to the maximum achievable flow of a network, e.g., $R^t$. Its average over one episode is calculated as $\frac{1}{2000}\sum_{t = 1}^{2000}{R^t}$.
\item \textbf{Loss}. This represents the TD-error of one training step.
\end{itemize}

Lastly, we note that as our problem is new, there are no other solutions we can compare against fairly. 
Hence, we only benchmark against the following alternative solutions:
\begin{itemize}
\item \textbf{Random (Rand) selection}: The UAV randomly selects a node or the base location as the destination for the next time slot.
\item \textbf{Maximum AoT First (MAF)}: The UAV always selects the device with the maximum AoT value $\delta^t_i$ as the destination for the next time slot.
\item \textbf{Nearest First (NF)}: The UAV selects the nearest device that has not been visited in the previous time slot.
\end{itemize}
\begin{table}[htbp]\centering
  \caption{Key system parameter values.}
  \label{simparameters}
  \begin{tabular}{ll}
    \midrule
    \textit{\textbf{Model Parameters}} & \textit{\textbf{Value}} \\
    \midrule
    Number of ground devices             & 7 \\
    Duration of one slot             & 300 s \\
    Region area                        & 2.5 $\times$ 2.5 km \\
    Number of slots $T$  & 2,000 \\
    Hovering blade profile power $P_{0}$~\cite{8663615} & 79.86 W \\
    Hovering induced power $P_{1}$~\cite{8663615} & 88.63 W \\
    Tip speed of the rotor blade $V_{0}$~\cite{8663615} & 120 m/s \\
    Average hovering rotor induced velocity $V_{1}$~\cite{8663615} & 4.03 m/s \\
    Fuselage drag ratio $\kappa$~\cite{8663615}        & 0.6 \\
    Air density $D_{\mathrm{air}}$~\cite{8663615}  & 1.225 kg/m$^3$ \\
    Rotor solidity $S_{rot}$~\cite{8663615}          & 0.05 \\
    Rotor disc area $A_{\mathrm{rot}}$~\cite{8663615} & 0.503 m$^2$ \\
    Maximum battery capacity $B_{\text{max}}$~\cite{DJIUAV} & 77 Wh \\
    Maximum UAV velocity ($V$)~\cite{DJIUAV}      & 21 m/s \\
    Maximum base location energy reservation $\hat{B}_\text{max}$      & 0.77 kWh \\
    Size of the solar panel $\Phi$      & 10 m$^2$ \\
    Solar energy conversion efficiency $\overline{\eta}$~\cite{7008488}      & 0.15 \\
    Network setup-1 (N1)       & 5 nodes, \\
                                   & 1.5 $\times$ 1.5 km  \\
    Network setup-2 (N2)       & 6 nodes, \\
                                   & 2 $\times$ 2 km  \\
    Network setup-3 (N3)       & 7 nodes, \\
                                   & 2.5 $\times$ 2.5 km  \\
    Network setup-4 (N4)       & 8 nodes, \\
                                   & 3 $\times$ 3 km  \\
                                   
    \multicolumn{2}{l}{} \\
    \midrule
    \textit{\textbf{Algorithm Parameters}} & \textit{\textbf{Value}} \\
    \midrule
    Number of episodes     & $100$ \\
    Learning rate $\eta$             & $10^{-4}$ \\
    Reward decay rate $\gamma$          & 0.5 \\
    Weight of AoT $\theta_1$        & 10 \\
    Weight of throughput $\theta_2$     & 0.5 \\
    Number of actions $N_A$            & 7 \\
    Size of the memory dataset         & 4,000 \\
    Mini-batch size                    & 32 \\
    Time interval to update $\theta'$ $U$ & Every 200 steps \\
    Neural network architecture       & Three fully \\
                                       & connected layers \\
    Activation function of neurons     & ReLU \\
    Optimizer     & Adam \\
    Priority exponent $\alpha$     & 0.2 \\
    Importance sampling exponent $\beta$ &  0.6 to 1 in training\\
    Priority bias $\epsilon_{P}$     & 10$^{-5}$ \\
    Exploration rate $\epsilon_{G}$ &  1 to 0.05  in training \\
    Training initiation threshold                & 320 transitions \\
    Assessment start episode              & 80th \\
    \hline
  \end{tabular}
\end{table}
%

%

\subsection{Results}
\begin{figure*}[htbp]
    \centering
    \subfloat[Reward vs. episode.]{
        \includegraphics[width=0.318\textwidth]{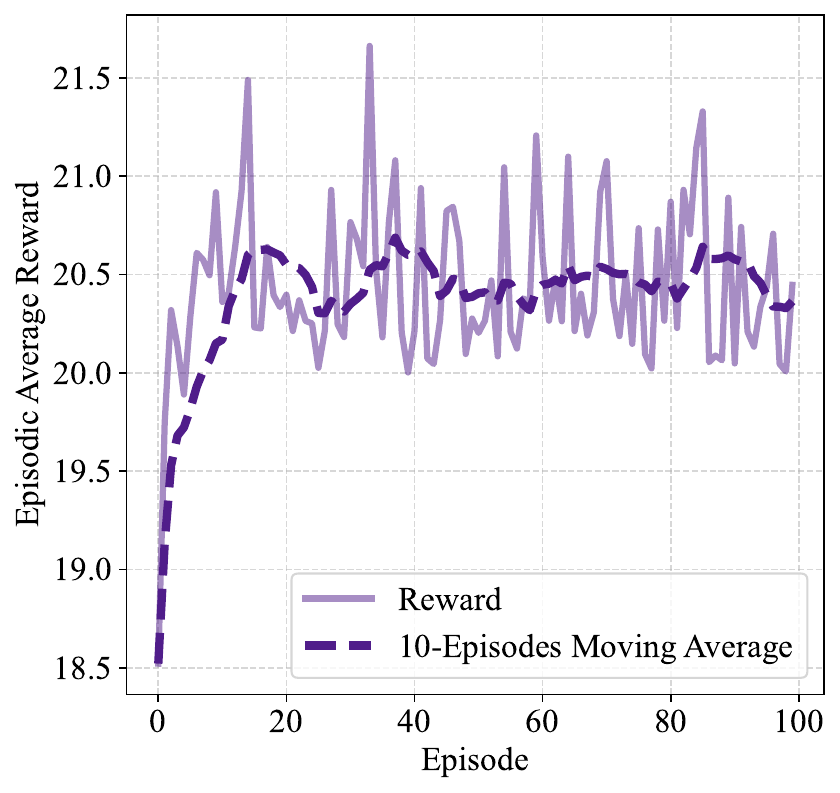}
        \label{ep_vs_reward_PD3QN}
    }
    \hfill
    \subfloat[AoT vs. episode.]{
        \includegraphics[width=0.31\textwidth]{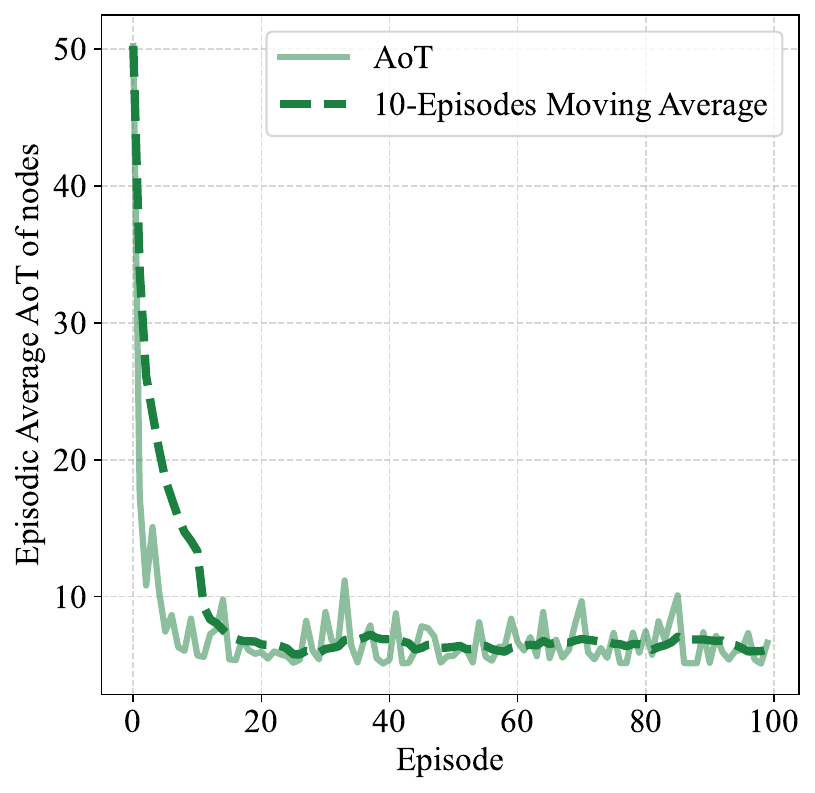}
        \label{ep_vs_delta_PD3QN}
    }
    \hfill
    \subfloat[Network throughput vs. episode.]{
        \includegraphics[width=0.31\textwidth]{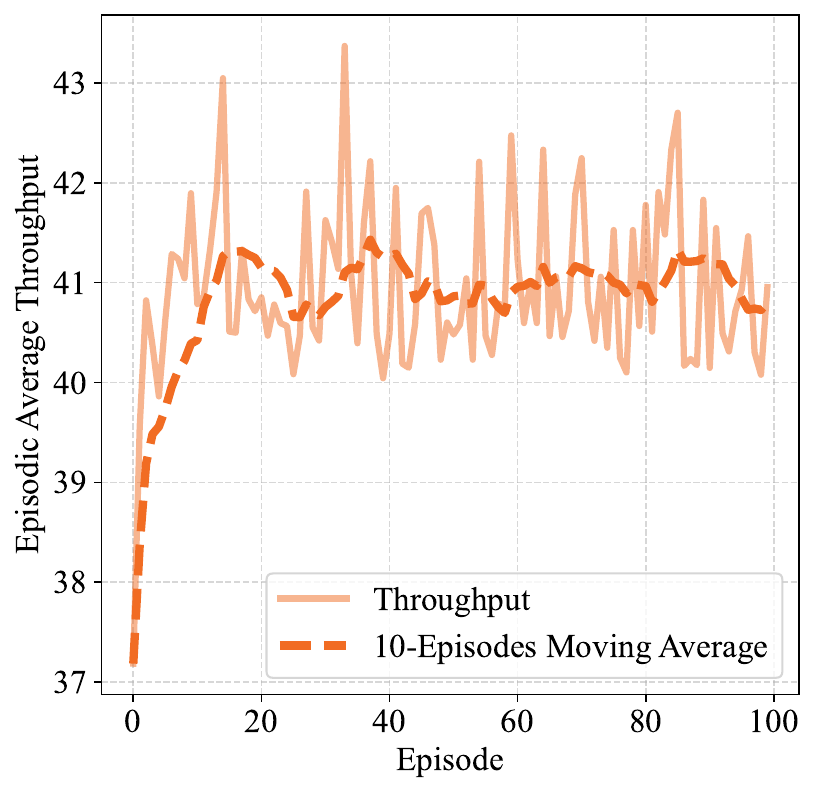}
        \label{ep_vs_flow_PD3QN}
    }

    \caption{Average reward, AoT value and network throughput over 100 episodes.}
    \label{ep_results_PD3QN}
\end{figure*}

Fig.~\ref{ep_results_PD3QN} presents the average reward gained by PD3QN, AoT of nodes and throughput achieved over 100 episodes (x-axis), using a 10-episode moving average to smooth out fluctuations. Fig.~\ref{ep_vs_reward_PD3QN} displays the progression of the average reward gained by PD3QN over one episode during training. It is evident that the UAV's reward increases over time, from 18.5 to 20.5, indicating that it has effectively learned from the environment and refined its strategy. Around the 20th episode, the reward reaches a plateau of around 20.5, suggesting that PD3QN has sufficiently explored the environment and converged. 
Furthermore, as shown in Fig.~\ref{ep_results_PD3QN}, even after convergence, the reward fluctuated between 20 and 21.5. This variation can be attributed to the randomness of solar energy arrivals at the base location, which directly impacted the UAV's energy levels. In ``Poor'' solar conditions, the UAV's flight range and mobility are constrained. Conversely, under favorable sunlight conditions, the UAV can operate longer without frequent returns for recharging, allowing it to complete tasks more effectively.

Fig.~\ref{ep_vs_delta_PD3QN} shows the evolution of the average AoT over the course of the training episodes. A clear decreasing trend can be observed, with AoT reducing from 50 to six, an 88\% reduction. This result suggests that after training, PD3QN has learned to reduce the AoT of a network.

Fig.~\ref{ep_vs_flow_PD3QN} records the average network throughput of a network over 100 episodes. The network throughput increased from 37.2 Kbps to approximately 41.1 Kbps; note, that the maximum network throughput is 50 Kbps.  This means PD3QN managed to reduce the loss in network throughput by 30\% after training.  Combining the results from Fig.~\ref{ep_vs_delta_PD3QN} and~\ref{ep_vs_flow_PD3QN}, we see that PD3QN successfully learned the best policy to optimize both AoT and network throughput simultaneously.
\begin{figure}[htbp]
	\centering 
	\includegraphics[scale = 0.56]{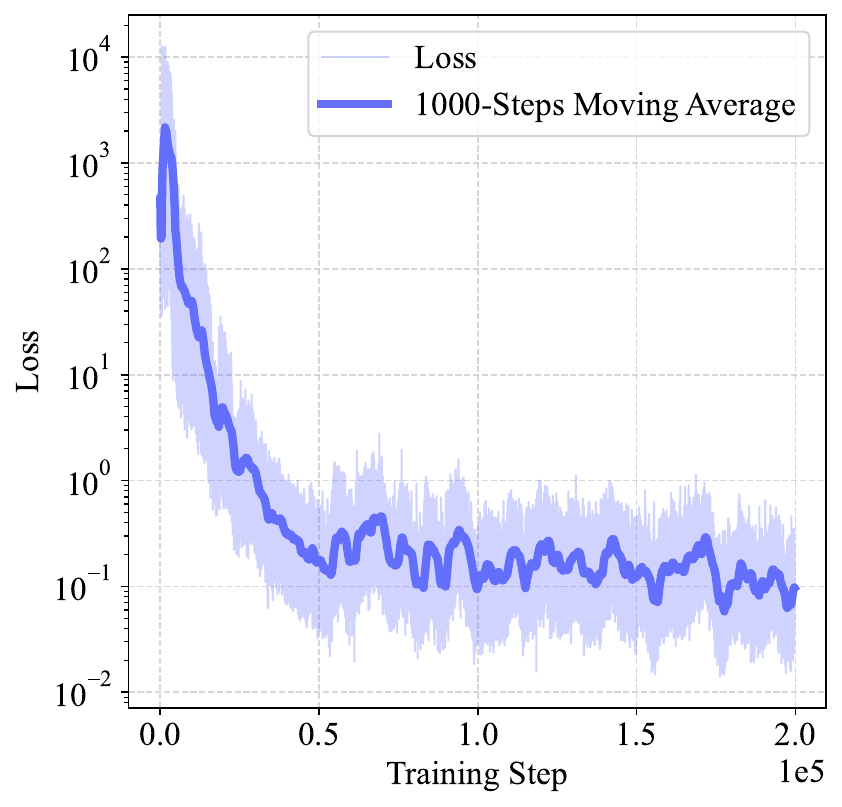}
	\caption{Loss versus training steps.}
	\label{step_vs_losses_PD3QN}
\end{figure}

Fig.~\ref{step_vs_losses_PD3QN} illustrates training loss behavior. The x-axis represents the number of training steps, while the y-axis displays the loss values on a logarithmic scale. A 1000-step moving average is applied to smooth fluctuations. 
We can see that the training loss shows a steep initial decline from 10$^4$ to around 10$^0$ within the first 50,000 training steps. This indicates that the UAV equipped with a PD3QN agent effectively learns and optimizes its parameters during this phase. Beyond this point, the training loss gradually decreased to 10$^{-1}$, meaning that the algorithm has reached convergence.
\begin{figure}[htbp]
	\centering 
	\includegraphics[scale = 0.56]{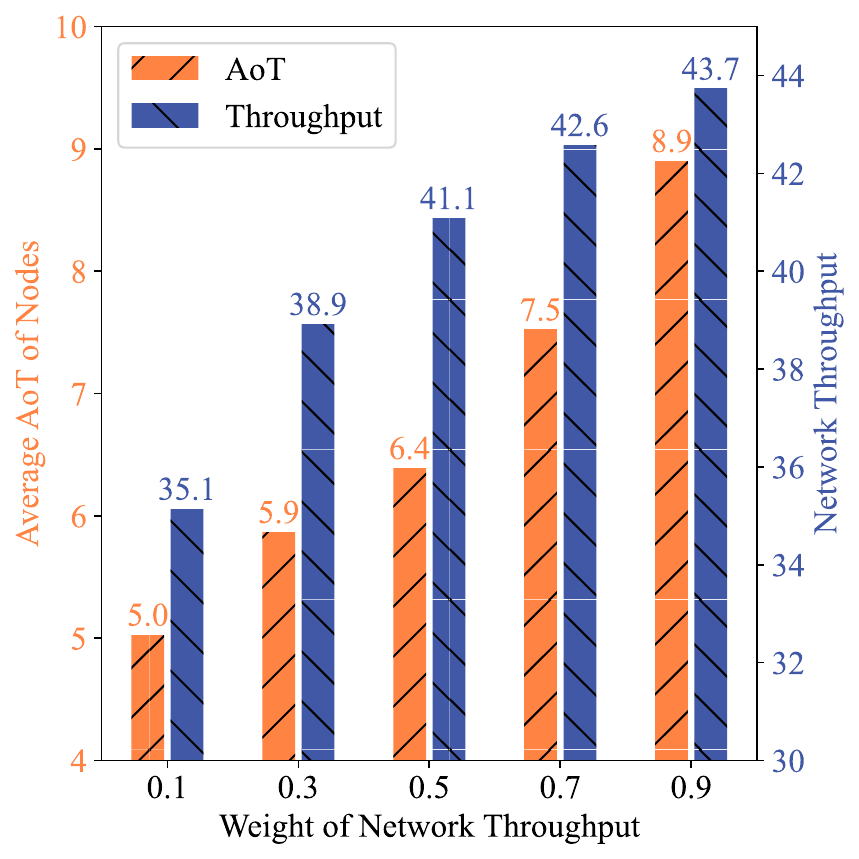}
	\caption{The objective weight $\theta_2$ versus AoT and network throughput.}
	\label{ratio_vs_delta_flow_PD3QN}
\end{figure}
Next, we evaluate how varying the objective weight $\theta_2$ affects the average AoT and network throughput.  We fixed $\theta_1=10$ and increased $\theta_2$ from 0.1 to 0.9 in steps of 0.2 to allow us to progressively emphasize network throughput in the reward.  We run 100 episodes to ensure statistical significance and calculated the mean for both AoT and network throughput.
Fig.~\ref{ratio_vs_delta_flow_PD3QN} shows the average AoT and network throughput for five different $\theta_2$ values (0.1, 0.3, 0.5, 0.7, 0.9).  AoT is represented in orange, while network throughput is shown in blue.
As depicted in Fig.~\ref{ratio_vs_delta_flow_PD3QN}, as $\theta_2$ increased from 0.1 to 0.9, AoT rose from 5.0 to 8.9, which indicates reduced node trustworthiness. Conversely, network throughput improved from 35.1 to 43.7 Kbps.
These results show the following trade-off: prioritizing network throughput leads to a reduction in trustworthiness. These findings suggest that by adjusting $\theta_2$, an operator can balance the optimization of network throughput against node trustworthiness, tailoring PD3QN to meet specific system requirements and priorities. For applications where data integrity and node reliability are paramount, a lower $\theta_2$ value should be chosen. Conversely, in scenarios prioritizing data flow, a higher $\theta_2$ value would be more appropriate.

\begin{figure}[htbp]
	\centering 
	\includegraphics[scale = 0.58]{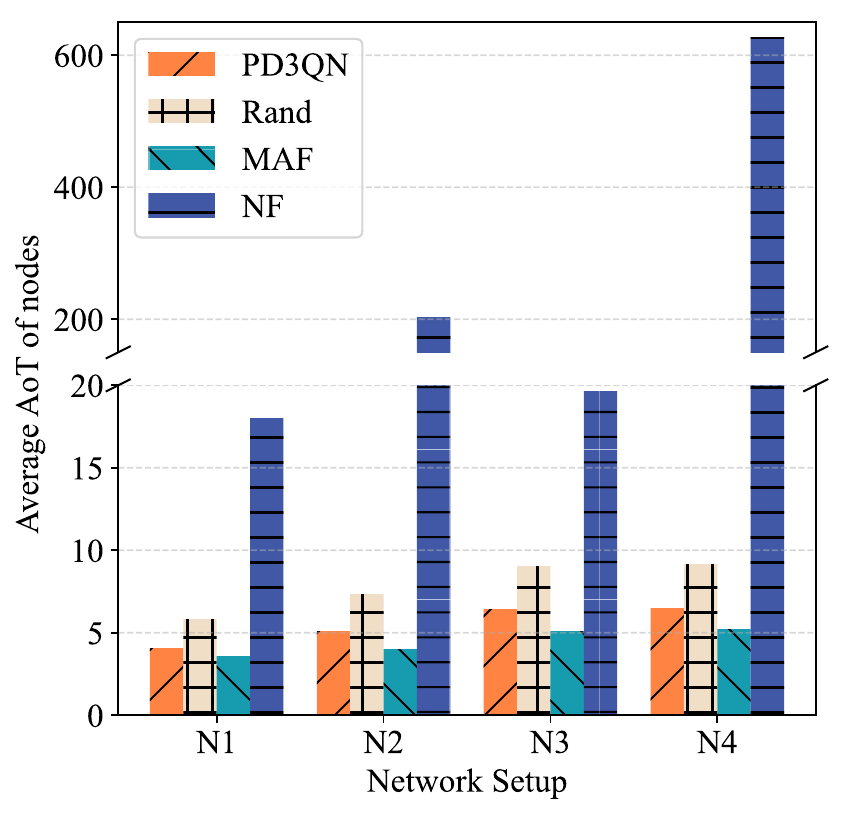}
	\caption{The AoT achieved by PD3QN and three competitive policies in four different network setups.}
	\label{delta_diff_setups}
\end{figure}

\begin{figure}[htbp]
	\centering 
	\includegraphics[scale = 0.56]{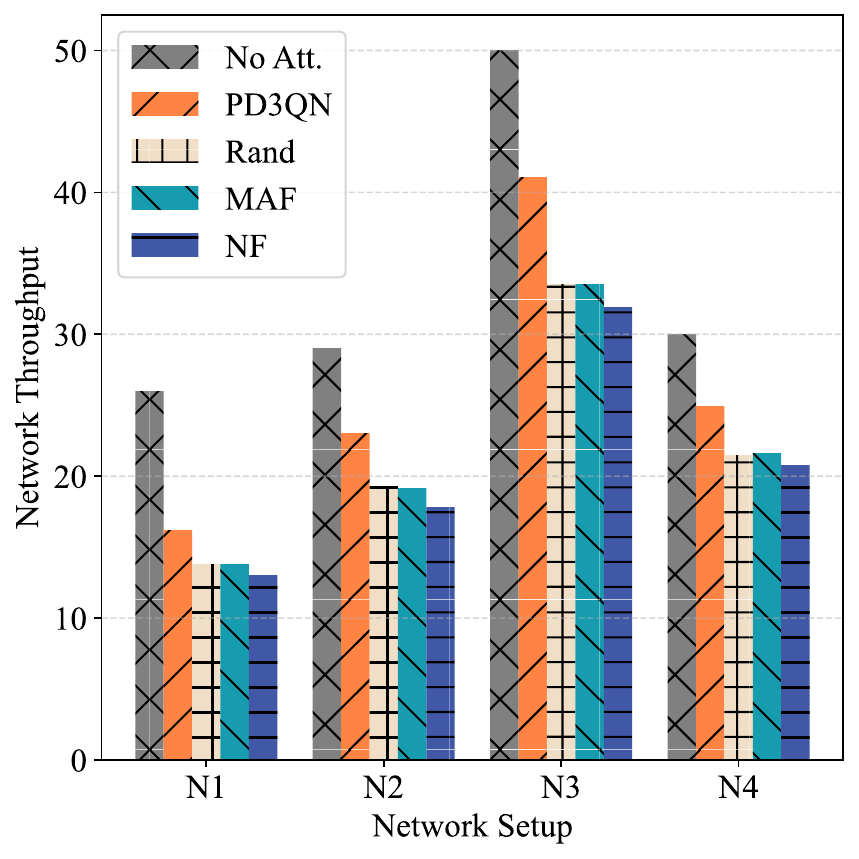}
	\caption{The network throughput achieved by PD3QN and three competitive policies and the throughput when there is no attestation in four different network setups.}
	\label{flow_diff_setups}
\end{figure}
Next, we evaluate the average AoT of devices and network throughput over 100 episodes to verify the effectiveness of PD3QN across four different network setups, labeled N1 through N4. 
Referring to Figure~\ref{delta_diff_setups}, MAF consistently achieved the lowest AoT, reaching 5.08 in the N3 setup, but at the cost of significantly lower throughput (see Fig.~\ref{flow_diff_setups}). This is because, under the MAF policy, the UAV always selected nodes with the highest AoT, effectively reducing the average AoT of devices. However, there is a trade-off between  AoT and throughput, where frequent device attestation may lower network throughput. As shown in Figure~\ref{flow_diff_setups}, the MAF policy ranked the second worst in terms of network throughput.
From Figure~\ref{delta_diff_setups}, PD3QN achieved the second-lowest AoT, with a value of 6.39 in the N3 scenario. Notably, in scenario N1 and N4, the AoT of PD3QN was close to that achieved by MAF. These results show that as compared to MAF, PD3QN managed to reduce not only the AoT of devices but also the reduction in throughput due to attestation.

On the other hand, from Figure~\ref{flow_diff_setups}, PD3QN consistently achieved the highest network throughput across all network setups. Specifically, in the N3 scenario, the network throughput achieved by PD3QN is 41.08 Kbps, which is closest to the maximum throughput of 50 Kbps when there is no attestation (No Att.), with a difference percentage of 17.8\%. Moreover, in the N3 scenario, PD3QN outperformed MAF, Rand and NF by 22.5\%, 22.6\% and 29.0\% in network throughput, respectively.
In contrast, Rand and NF achieved a higher AoT and lower network throughput compared to PD3QN. Specifically, in the N3 setup, Rand attained an AoT of 9.0, significantly higher than that of PD3QN (6.39) and MAF (5.08). The NF policy consistently performed worst, yielding both the highest AoT and lowest throughput across all setups. %
Both Figure~\ref{delta_diff_setups} and~\ref{flow_diff_setups} indicate that PD3QN was able to effectively balance the trade-off between minimizing AoT and maximizing network throughput across various network configurations, outperforming traditional policies in both key performance metrics.
%

\section{Conclusion} \label{sec:conc}
This paper has considered a novel approach for device attestation in IoT networks, combining AoT, UAVs, and reinforcement learning. The proposed method, PD3QN, optimizes both device AoT and network throughput and effectively adapts to the time-varying energy supply at a charging station.

Simulation results demonstrate that, after training, PD3QN reduces AoT from approximately 50 to 6, achieving an 88\% reduction, while increasing network throughput from 37.2 to 41.0 Kbps. Compared to the maximum achievable network throughput of 50 Kbps without attestation, PD3QN reduces throughput loss by 30\%. PD3QN effectively balances trustworthiness and throughput.

In future research, we will address the "curse of dimensionality" in MDPs as the number of devices grows, particularly regarding the AoT vector. We will apply state aggregation, clustering devices located in the same region into virtual nodes. This simplifies computations and reduces the state space, improving DRL algorithm scalability for larger IoT deployments. Building upon this approach, we will expand our solution to multi-UAV and multi-charging-station collaborative scenarios. In these scenarios, state aggregation will help effectively manage the increased complexity. Additionally, we will explore Multi-Agent Reinforcement Learning (MARL) approaches to enable effective cooperation among multiple UAVs for IoT node verification. With a reduced state space enabled by aggregation, MARL will support faster training and provide enhanced scalability and security for large-scale IoT networks.
%

%
\bibliographystyle{IEEEtran}
\bibliography{Refs.bib}

\begin{thebibliography}{10}
\providecommand{\url}[1]{#1}
\csname url@samestyle\endcsname
\providecommand{\newblock}{\relax}
\providecommand{\bibinfo}[2]{#2}
\providecommand{\BIBentrySTDinterwordspacing}{\spaceskip=0pt\relax}
\providecommand{\BIBentryALTinterwordstretchfactor}{4}
\providecommand{\BIBentryALTinterwordspacing}{\spaceskip=\fontdimen2\font plus
\BIBentryALTinterwordstretchfactor\fontdimen3\font minus \fontdimen4\font\relax}
\providecommand{\BIBforeignlanguage}[2]{{%
\expandafter\ifx\csname l@#1\endcsname\relax
\typeout{** WARNING: IEEEtran.bst: No hyphenation pattern has been}%
\typeout{** loaded for the language `#1'. Using the pattern for}%
\typeout{** the default language instead.}%
\else
\language=\csname l@#1\endcsname
\fi
#2}}
\providecommand{\BIBdecl}{\relax}
\BIBdecl

\bibitem{6198335}
M.~Cheminod, L.~Durante, and A.~Valenzano, ``Review of security issues in industrial networks,'' \emph{IEEE Trans. on Ind. Inform.}, vol.~9, no.~1, pp. 277--293, Feb. 2013.

\bibitem{8027075}
J.~Wang, Z.~Hong, Y.~Zhang, and Y.~Jin, ``Enabling security-enhanced attestation with intel {SGX} for remote terminal and {IoT},'' \emph{IEEE Trans. Comput.-Aided Design Integr. Circuits Syst.}, vol.~37, no.~1, pp. 88--96, Jan. 2018.

\bibitem{9096052}
M.~Ammar, B.~Crispo, and G.~Tsudik, ``Simple: A remote attestation approach for resource-constrained {IoT} devices,'' in \emph{ACM/IEEE International Conference on Cyber-Physical Systems (ICCPS)}, Sydney, NSW, Australia, Apr. 2020, pp. 247--258.

\bibitem{1301329}
A.~Seshadri, A.~Perrig, L.~van Doorn, and P.~Khosla, ``{SWATT:} software-based attestation for embedded devices,'' in \emph{IEEE Symposium on Security and Privacy}, Berkeley, CA, USA, May 2004, pp. 272--282.

\bibitem{10.1145/2508859.2516650}
F.~Armknecht, A.-R. Sadeghi, S.~Schulz, and C.~Wachsmann, ``A security framework for the analysis and design of software attestation,'' in \emph{ACM SIGSAC Conf. on Computer and Commun. Security}, Berlin, Germany, Nov. 2013.

\bibitem{SEDA}
N.~Asokan, F.~Brasser, A.~Ibrahim, A.-R. Sadeghi, M.~Schunter, G.~Tsudik, and C.~Wachsmann, ``{SEDA}: Scalable embedded device attestation,'' in \emph{Proc. of the 22nd ACM SIGSAC Conference on Computer and Communications Security}, Denver, Colorado, USA, 2015, p. 964–975.

\bibitem{TPMbook}
W.~Arthur and D.~Challener, \emph{A Practical Guide to {TPM} 2.0: Using the Trusted Platform Module in the New Age of Security}, 1st~ed.\hskip 1em plus 0.5em minus 0.4em\relax USA: Apress, Jan. 2015.

\bibitem{xiao2024AoT}
\BIBentryALTinterwordspacing
Y.~Xiao, Q.~Du, W.~Cheng, P.~D. Diamantoulakis, and G.~K. Karagiannidis, ``Age of trust {(AoT)}: A continuous verification framework for wireless networks,'' 2024. [Online]. Available: \url{https://arxiv.org/abs/2406.02190}
\BIBentrySTDinterwordspacing

\bibitem{10869306}
J.~Pan, Y.~Li, R.~Chai, S.~Xia, and L.~Zuo, ``Multi-objective trajectory planning for {UAV}-assisted {IoT} networks based on {DRL} approach,'' \emph{IEEE Internet Things J.}, pp. 1--1, Jan. 2025.

\bibitem{10820863}
G.~Zhang, X.~Wei, X.~Tan, Z.~Han, and G.~Zhang, ``{AoI} minimization based on deep reinforcement learning and matching game for {IoT} information collection in {SAGIN},'' \emph{IEEE Trans. Commun.}, pp. 1--1, Jan. 2025.

\bibitem{10836723}
Y.~Liu, Q.~Deng, Z.~Zeng, A.~Liu, and Z.~Li, ``A hybrid optimization framework for age of information minimization in {UAV}-assisted {MCS},'' \emph{IEEE Trans. Serv. Comput.}, pp. 1--17, Jan. 2025.

\bibitem{10877810}
S.~Bai, X.~Wang, M.~Cenk~Gursoy, G.~Jiang, and S.~Xu, ``Deep reinforcement learning for rechargeable {UAV}-assisted data collection from dense mobile sensor nodes,'' \emph{IEEE Access}, pp. 1--1, Feb. 2025.

\bibitem{10838612}
A.~Singh and R.~M. Hegde, ``Age-aware {UAV}-aided energy harvesting for the design of wireless rechargeable mobile networks,'' \emph{IEEE Trans. Artif. Intell.}, pp. 1--11, Jan. 2025.

\bibitem{10759639}
X.~Peng, X.~Lan, and Q.~Chen, ``Age of task-aware {UAV}-based mobile edge computing techniques in emergency rescue applications,'' \emph{IEEE Internet Things J.}, pp. 1--1, Nov. 2024.

\bibitem{10188891}
W.~Liu, D.~Li, T.~Liang, T.~Zhang, Z.~Lin, and N.~Al-Dhahir, ``Joint trajectory and scheduling optimization for age of synchronization minimization in {UAV}-assisted networks with random updates,'' \emph{IEEE Trans. Commun.}, vol.~71, no.~11, pp. 6633--6646, Jul. 2023.

\bibitem{9562134}
Z.~Cui, T.~Yang, X.~Wu, C.~Li, C.~Wang, and B.~Hu, ``The learning stimulated sensing-transmission coordination via age of updates in distributed {UAV} swarm,'' in \emph{Intl. Symp. on Wirel. Commun. Sys.}, Berlin, Germany, Sep. 2021, pp. 1--6.

\bibitem{10858287}
M.~Hao, C.~Shang, S.~Wang, W.~Jiang, and J.~Nie, ``{UAV}-assisted zero knowledge model proof for generative {AI}: A multi-agent deep reinforcement learning approach,'' \emph{IEEE Internet Things J.}, pp. 1--1, Jan. 2025.

\bibitem{swA2016}
R.~V. Steiner and E.~Lupu, ``Attestation in wireless sensor networks: A survey,'' \emph{ACM Computing Surveys (CSUR)}, vol.~49, no.~3, Sep. 2016.

\bibitem{JointRLinkS}
M.~Kodialam and T.~Nandagopal, ``Characterizing achievable rates in multi-hop wireless networks: the joint routing and scheduling problem,'' in \emph{ACM Mobicom}, San Diego, CA, USA, 2003, p. 42–54.

\bibitem{ChannelRLinkS}
M.~Alicherry, R.~Bhatia, and L.~E. Li, ``Joint channel assignment and routing for throughput optimization in multi-radio wireless mesh networks,'' in \emph{ACM Mobicom}, Cologne, Germany, 2005, p. 58–72.

\bibitem{8663615}
Y.~Zeng, J.~Xu, and R.~Zhang, ``Energy minimization for wireless communication with rotary-wing {UAV},'' \emph{IEEE Trans. on Wirel. Commun.}, vol.~18, no.~4, pp. 2329--2345, Apr. 2019.

\bibitem{7008488}
M.~Ku, Y.~Chen, and K.~J.~R. Liu, ``Data-driven stochastic models and policies for energy harvesting sensor communications,'' \emph{IEEE J. Sel. Areas Commun.}, vol.~33, no.~8, pp. 1505--1520, Aug. 2015.

\bibitem{BELLMAN1957MDP}
R.~Bellman, ``A markovian decision process,'' \emph{Journal of Mathematics and Mechanics}, vol.~6, no.~5, pp. 679--684, 1957.

\bibitem{Mnih2015HumanlevelCT}
V.~Mnih, K.~Kavukcuoglu, D.~Silver, A.~A. Rusu, J.~Veness, M.~G. Bellemare, A.~Graves, M.~A. Riedmiller, A.~Fidjeland, G.~Ostrovski, S.~Petersen, C.~Beattie, A.~Sadik, I.~Antonoglou, H.~King, D.~Kumaran, D.~Wierstra, S.~Legg, and D.~Hassabis, ``Human-level control through deep reinforcement learning,'' \emph{Nature}, vol. 518, pp. 529--533, 2015.

\bibitem{double_Q}
H.~v. Hasselt, A.~Guez, and D.~Silver, ``Deep reinforcement learning with double {Q}-learning,'' in \emph{AAAI}, Phoenix, Arizona, Feb. 2016, p. 2094–2100.

\bibitem{DuelingPaper}
Z.~Wang, T.~Schaul, M.~Hessel, H.~Hasselt, M.~Lanctot, and N.~Freitas, ``Dueling network architectures for deep reinforcement learning,'' in \emph{Intl. Conf. on Machine Learning}, vol.~48, New York, New York, USA, Jun. 2016, pp. 1995--2003.

\bibitem{sutton2018RLBOOK}
R.~S. Sutton and A.~G. Barto, \emph{Reinforcement learning: An introduction}.\hskip 1em plus 0.5em minus 0.4em\relax Cambridge, USA: MIT press, 2018.

\bibitem{schaul2016prioritizedexperiencereplay}
\BIBentryALTinterwordspacing
T.~Schaul, J.~Quan, I.~Antonoglou, and D.~Silver, ``Prioritized experience replay,'' Feb. 2016. [Online]. Available: \url{https://arxiv.org/abs/1511.05952}
\BIBentrySTDinterwordspacing

\bibitem{DJIUAV}
\BIBentryALTinterwordspacing
{DJI} {Mavic} 3. [Online]. Available: \url{https://www.dji.com/support/product/mavic-3}
\BIBentrySTDinterwordspacing

\end{thebibliography}
\end{document}